# Optomechanical sensor network with fiber Bragg gratings

Shiwei Yang[1*], Qiang Zhang[1,2*†], Linrun Yang[1], Hanghua Liu[1], Quansen Wang[1], Pengfei Zhang[1,2], Heng Shen[1,2†], Yongmin Li[1,2,3†]

[1]State Key Laboratory of Quantum Optics and Quantum Optics Devices, Institute of Opto-Electronics, Shanxi University, Taiyuan 030006, China

[2]Collaborative Innovation Center of Extreme Optics, Shanxi University, Taiyuan 030006, China

[3]Hefei National Laboratory, Hefei 230088, China.

*These authors contributed equally to this work.

†Corresponding author: qzhang@sxu.edu.cn, hengshen@nbi.dk, and yongmin@sxu.edu.cn

Cavity optomechanics offers a versatile platform for both fundamental physics and ultrasensitive sensing. Importantly, resonant enhancement in both optical and mechanical responses enables the highly sensitive optical detection of small forces, displacements, vibrations, and magnetic fields, enabling it a promising candidate of the next generation of ultrasensitive sensor networks. However, this is impeded by the fiber optic-incompatibility and intrinsic nature of existing optomechanical sensors. Here, we report the first demonstration of an optomechanical sensor network in terms of magnetic field detection, wherein multiple fiber-optic optomechanical sensors are connected into a standard single mode fiber. Building upon a commercially available fiber Bragg gratings, we realize a robust low-loss, low-noise, and polarization-insensitive coupling with light sources in a way compatible with fiber optics. This thus enables our optomechanical senor to fulfill the requirements for ultrasensitive sensor networks. Furthermore, in this sensor network we demonstrate the sensitivity of 8.73 pm/Gs for DC magnetic fields and 537 fT/Hz$^{1/2}$ for AC magnetic fields in a magnetically unshielded environment with the ambient temperature and pressure, better than the reported values in previous optomechanical magnetometers. Our work sheds light on exploiting cavity optomechanics in the practical applications and ultrasensitive senor networks.

Sensor networks exploit multiple sensors distributed across the area of interest to obtain the comprehensive information of the targets, playing essential roles in a variety of applications ranging from internet of things[1], smart cities[2], oil and gas industry[3], seafloor faults and ocean dynamics[4], to earthquake and volcano dynamic detection[5,6]. Of particular importance is the key building block of the network, the sensor, which should fulfill the requirement of high-sensitivity, robustness, scalability, and fiber optics compatibility for multiplex.

In recent year, cavity optomechanics emerges and offers many ultrasensitive single-point sensing mechanisms and sensors[7–43]. Benefiting from high-Q optical and mechanical resonances, cavity optomechanical sensors have realized ultrasensitive optical detection of small forces[15,16], displacements[17–20], rotations[21–23], vibrations[24–28], ultrasounds[29–31], radio waves[32–36], magnetic fields[37–41], and other quantities[42,43]. However, it is still an open challenge to build an optomechanical sensor network for those existing optomechanical sensors. Multiple sensors based on Fabry-Perot cavities or levitated particles cannot be cascaded and integrated into a standard single mode fiber (SMF). On the other hand, evanescent fields coupling for whispering gallery mode (WGM) and photonic crystal optomechanical systems introduces massive low-frequency thermal noises from tapered fibers or excess loss from on-chip waveguides[44,45], and polarization variability from random birefringence in SMF especially for long-distance networks[46] can result in time-variant coupling between the probe light and the designated optical mode. In this regard, a robust optomechanical systems with low-noise and low-loss is highly demanded for building an optomechanical sensor network.

In this work, we demonstrate an optomechanical sensor network with multiple fiber-optic optomechanical sensors connected into a standard SMF. Our optomechanical sensor consists of a fiber-optic mechanical resonator (MR) incorporating a phase-shifted fiber Bragg grating (PFBG) and a functional support. A minute perturbation excites the fiber MR and changes the period of PFBG, and thus can be detected by measuring the reflective light from the PFBG with high-$Q$ optical resonances. Using the optical fiber MR and PFBG, a robust low-loss, low-noise, and polarization-insensitive coupling with light sources is realized, eliminating the limitations for previous optomechanical sensors based on evanescent fields coupling.

Furthermore, as a proof-of-concept we apply this optomechanical sensor network for the optical detection of magnetic fields. By attaching a magnetostrictive material (Terfenol-D) in response to the ambient magnetic field, the proposed sensor acts as a fiber-optic optomechanical magnetometer (FOMM). We demonstrate that the FOMM enables the detection of both DC and AC magnetic fields in magnetically unshielded environment with ambient temperature and pressure. A peak DC magnetic-field sensitivity of 8.73 pm/Gs (the resonant wavelength shift of the PFBG per unit Gauss) is achieved, slightly better than the best value of 7.57 pm/Gs with WGM[47]. Meanwhile, the peak AC magnetic-field sensitivity of 537 fTHz$^{-1/2}$, is 48 times better than what can be reached by existing optomechanical magnetometers (26 pTHz$^{-1/2}$ in Ref. 40). Meanwhile, this device is only sensitive to the magnetic field along the fiber as a result of the Terfenol-D rod magnetized uniaxially, illustrating the potential of detecting three-dimensional vector magnetic fields using three orthogonal FOMMs. In addition, the resonant frequency of the Terfenol-D MR could be tuned by a bias magnetic field, enabling broadband measurement with the peak sensitivity.

**Results**
**Principle of operation**
Fig. 1a illustrates the schematic of an optomechanical sensor network, where multiple sensors are integrated into a standard SMF for quasi-distributed sensing. In this architecture, both DC and AC magnetic field can be detected. Concretely, we obtain the response to DC magnetic field by monitoring the reflective spectrum of the sensor network, wherein a broadband source (BBS) and an optical spectrum analyzer (OSA) are used. On the other hand, in order to probe the AC magnetic field, lights with different wavelengths are coupled into the sensor network by a wavelength division multiplexer (WDM) and a fiber circulator. We stress that this system works in a magnetically unshielded environment with the ambient temperature and pressure.

As the key element, the optomechanical sensor consists of a clamped-clamped fiber MR (blue) and a functional support (black) as shown in Fig.1b and c. The fiber MR is fabricated by packaging a standard PFBG between two silica capillaries. The abrupt junction between the PFBG and silica capillaries induces an acoustical mismatched impedance, confining the phonons into the PFBG to form a high-$Q$ mechanical resonator ($Q_m > 10^5$)[48]. Fig. 1d presents the power spectral densities (PSD) and vibrating modeshapes of the fundamental and third modes for a fiber MR with a length of 8 mm. An optical reflective spectrum of the sensor network with six sensors is shown in Fig. 1e, and each spectrum has a narrow dip, similar to the transmission spectrum of a WGM cavity.

As a demonstration of detecting the magnetic field, we further use a cuboid Terfenol-D rod as the functional support (in Fig. 1b and c) to response ambient magnetic field. Consequently, when an ambient magnetic field is applied to the FOMM, the Terfenol-D rod generates the associated deformation and thus changes the length and grating period of the fiber MR. By interrogating the reflection of the PFBG determined by the grating period, the ambient magnetic field could be measured in real time. Intriguingly, a large optomechanical coupling coefficient is promised by the coaxial coupling between motion of the Terfenol-D rod and the PFBG.

To improve the sensitivity, a mechanism of deformation transfer is used to amplify the strain of the fiber MR from the deformation of the Terfenol-D rod with a deformation transfer radio ($D_{TR}$) of $D_{TR} = L_1/L_2$.

Here the deformation of the Terfenol-D rod with a length of $L_1$ is transferred to fiber MR with a length of $L_2$, as shown in Fig. 1b.

Building upon such scalable and fiber-compatible magnetometers, a sensor network is constructed by simply injecting probe lights into the sensor network through a fiber-optic circulator. This thus provides a simple, robust, low-noise and low-loss coupling approach, overcoming the limitation of the evanescent fields coupling for the WGM and photonic crystal optomechanical systems. Meanwhile, the PFBG is isotropic for the polarizations of probe light, suppressing the influence of optical polarization fluctuations from random birefringence in SMF, especially for long-distance networks. Furthermore, using commercially available fiber Bragg gratings and WDMs promise a practical pre-distributed optomechanical sensing networks.

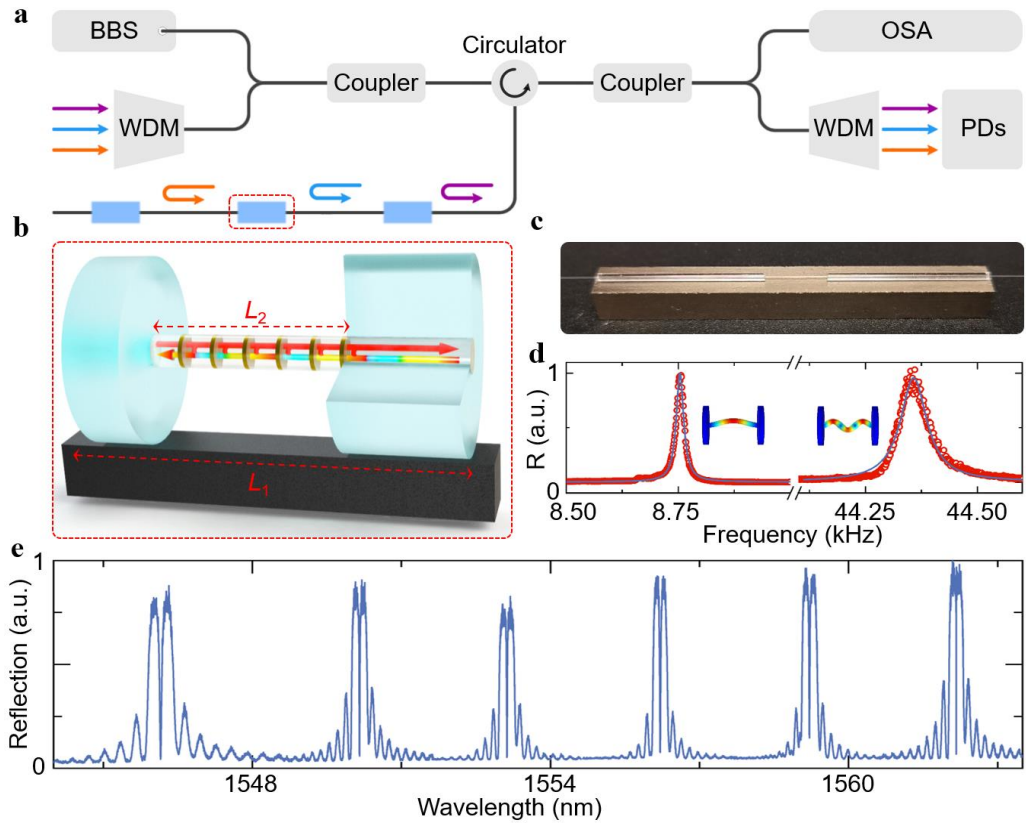

**Fig. 1. Optomechanical sensor network. a**, Experimental apparatus of the sensor network. **b**, Schematic diagram of the FOMM. **c**, Photograph of the FOMM. **d**, Power spectral density (PSD) of the fiber MR. Inset: Finite-element simulations of vibrating modeshapes of the fiber MR. **e**, Reflective spectrum of the sensor network with six PFBGs. BBS, broadband source; OSA, optical spectrum analyzer; WDM, wavelength division multiplexer; PDs, photodetectors.

**Response to DC magnetic fields**

We first investigate the system response to the DC magnetic fields. A broadband light source and an optical spectrum analyzer are used to measure the reflective spectrum shifts of the FOMMs with different $D_{TR}$ as shown in Fig. 1a. The resonant wavelength shift $\Delta\lambda$ resulted from the applied DC magnetic field $B$ is given by[49]

$$\Delta\lambda(B) = 0.79\lambda_c \alpha B D_{TR} \qquad (1)$$

where $\lambda_c$ is the dip wavelength of the PFBG in absence of the magnetic fields and $\alpha$ is magnetostrictive coefficient of the Terfenol-D rod.

In our experiment, the DC magnetic field is generated using an 18-mm-diameter coil, which is calibrated with a commercial magnetic sensor. Fig. 2 shows the reflective spectra of the FOMMs with different $D_{TR}$ under varying DC magnetic fields. We observe that the resonant dip of the PFBG shifts to the long wavelength region (redshift) as the DC magnetic fields increase. The wavelength shift is proportional to the $D_{TR}$ and ambient magnetic fields. The peak sensitivity is 8.73 pm/Gs for the FOMM with an $D_{TR}$ of 13.5, which is in good agreement with the theoretic values ($\lambda_c$ = 1550 nm and $\alpha$ = 0.528 ppm/Gs). The corresponding optomechanical coupling constant $g_{OM}$ is $2\pi \times 38.7$ MHz/nm, where $g_{OM} = \partial\omega_c/\partial x$ is defined as the resonant frequency shift per unit displacement.

In parallel, direction of magnetic field is another key parameter for measuring the ambient magnetic fields. In Fig. 2b, the greed curve represents the response of the FOMM ($D_{TR}$=13.5) to a magnetic field orthogonal to the axis of the Terfenol-D rod. The corresponding sensitivity is 0.01 pm/Gs, which is about one thousandth of that of the FOMM to a magnetic field coaxial with the Terfenol-D rod. Therefore, the FOMM realizes one-dimensional directional magnetic-field measurement, overcoming the restrict of direction-insensitivity in reported optomechanical magnetometers and providing a potential way for detecting three-dimensional vector magnetic fields.

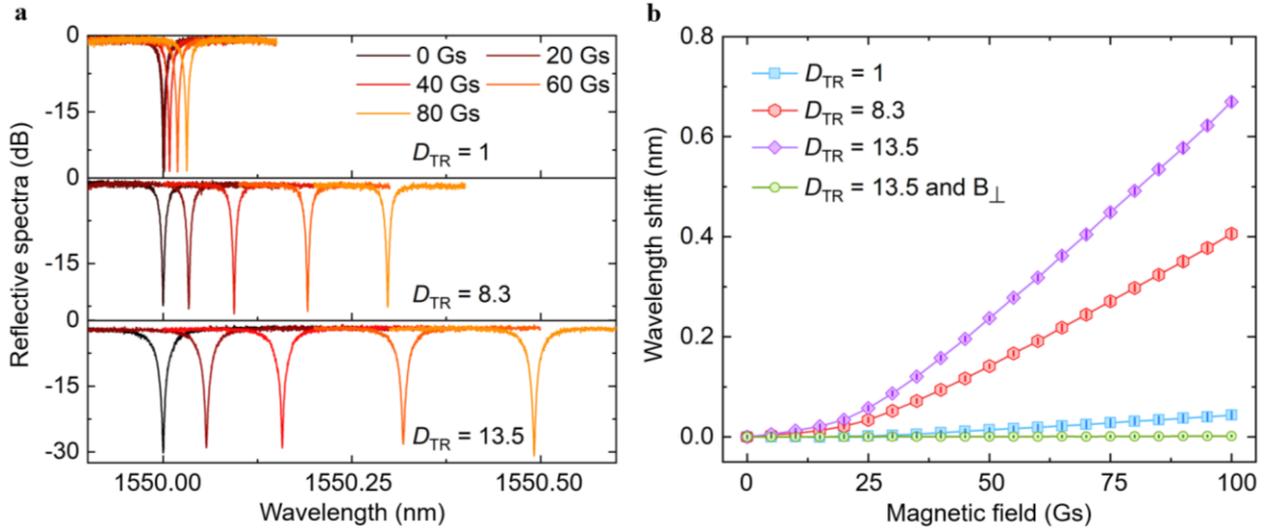

**Fig. 2. Response of the FOMMs to DC magnetic fields. a**, Reflective spectrum shifts of FOMMs with different $D_{TR}$ of 1, 8.3, and 13.5 as a function of DC magnetic fields. **b**, Resonant wavelength shifts of the three FOMMs as a function of coaxial and orthogonal DC magnetic fields.

**Response to AC magnetic fields**
In order to examine the response of the FOMM to AC magnetic fields, a probe light from a tunable laser is coupled into the FOMM to monitor the deformation of the fiber MR using the dispersive coupling mechanism, where the wavelength of the probe light is slightly detuned from the optical resonant dip. The reflective optical power with AC magnetic-field excitations is given by

$$P(\omega) = \eta P_{in} + \Delta\lambda k \quad (2)$$

where $\eta$ and $P_{in}$ are the reflectivity of the PFBG at the wavelength of the probe light and the input power of the probe light, and $k$ is the slope of the reflective spectrum of the PFBG and is achieved from the derivation of Lorentzian function

$$k(\Delta) = \frac{8\omega_{PFBG}^2 P_0 (R_{max} - R_{min})}{(4\Delta^2 + \omega_{PFBG}^2)^2} \Delta \quad (3)$$

where $\Delta = \lambda_p - \lambda_c$ is the laser-cavity detuning, $\omega_{PFBG}$ is the FWHM of the PFBG, $P_0$ is light power, $R_{max}$ is the maximum reflectivity of the PFBG, and $R_{min}$ is the reflectivity of the PFBG at optical resonant dip. When the frequency of the probe light is detuned by $\Delta = \pm\omega_{PFBG}/12^{1/2}$, the slope $k$ is maximized

$$k\left(\Delta = \pm\omega_{PFBG}/2\sqrt{3}\right) = \frac{3\sqrt{3}}{4}\frac{P_0(R_{max} - R_{min})}{\omega_{PFBG}} \quad (4)$$

The Response $R(\omega)$ in unit of dBm under AC magnetic excitation could be expressed as[25]

$$R(\omega) = 10\log\left(\frac{(g\gamma\Delta\lambda k)^2}{Z}1000\right) \quad (5)$$

where $g$ and $\gamma$ are the conversion gain and rate of the photodetector, and $Z$ is 50 Ω. For AC magnetic-field measurement, the mechanical resonances need to be considered, and the resonant wavelength shift $\Delta\lambda = 0.79\lambda_p\alpha D_{TR}\omega_{TD}^2\chi B(\omega)$, where $\omega_{TD}$ and $\chi(\omega)$ are the resonant frequency and mechanical susceptibility of the Terfenol-D MR. Finally, the AC magnetic-field sensitivity could be obtained as[38]

$$B_{min}(\omega) = \frac{B_{ref}}{\sqrt{SNR(\omega)\times RBW}} \quad (6)$$

where the signal-to-noise ratio (SNR) is the ratio of the response $R(\omega)$ to the noise floor.

Fig. 3 depicts a typical measurement at ambient temperature and pressure. Concretely, Fig. 3a shows the PSD of the noise floor (upper panel) and the response $R(\omega)$ (middle panel) of the FOMM with a $D_{TR}$ of 5.1, where the amplitude of AC magnetic-field $B_{ref}$ keeps constant at 24 nT, and the resolution bandwidth (RBW) of the spectrum analyzer is 10 Hz. The response peak in Fig. 3a (middle panel) locates at the position of the fundamental stretching mode of the Terfenol-D rod with a mechanical quality factor of 180. Experimental results show that the peak AC sensitivity is 537 fTHz$^{-1/2}$ at 11.188 kHz, as shown in Fig. 3a (bottom panel). According to the PSD in Fig. 3a, it is suggested that the thermal noise of the stretching eigenmode of the Terfenol-D MR is beneath the noise floor, indicating that the precision of this device could be improved by decreasing the noise floor.

The response range of the FOMM is evaluated by varying the amplitude of the applied AC magnetic field. Fig. 3b shows the normalized response of the FOMM at the resonant frequency (11.188 kHz) as a function of magnetic-field strength with 10 Hz resolution bandwidth. A linear scale is observed when the magnetic-field strength is less than 200 nT, demonstrating a linear range from 537 fT to 200 nT. Beyond 200 nT, the wavelength of the probe light is far away from the optical resonant dip and the linear slope region, and the system response reaches the flat regions of the reflective spectra of the PFBG. Thus, the response becomes constant afterwards.

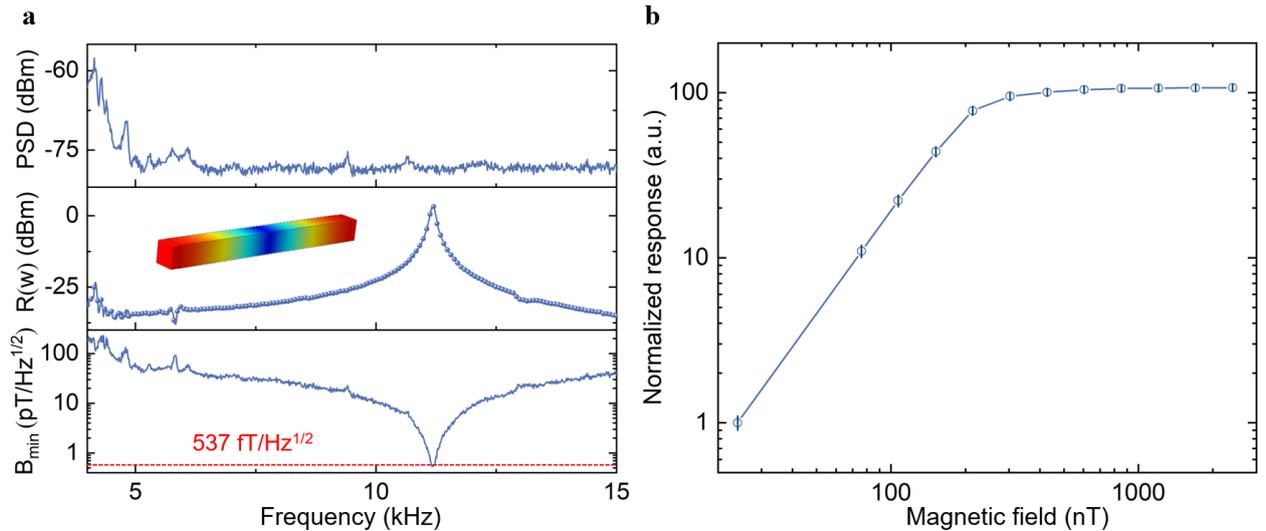

**Fig. 3. Response to AC magnetic fields. a**, PSD of the FOMM without magnetic excitation (upper), response $R(\omega)$ to frequency scanning magnetic excitation (middle), and resolution $B_{min}$ as a function of frequency (bottom). Inset: Finite-element modeling of mechanical resonant eigenmode of the Terfenol-D rod. **b**, Normalized response of the FOMM at the resonant frequency as a function of magnetic-field strength with 10 Hz resolution bandwidth.

In practice, broadband ultrasensitive measurement is challenging for the reported optomechanical magnetometers. However, our device addresses this issue by employing a bias magnetic field to adjust the resonant frequency of the Terfenol-D MR. For the FOMM with a $D_{TR}$ of 9.3, when the bias magnetic field increases from 0 Gs to 200 Gs, the resonant frequency of the Terfenol-D MR decreases about 1.3 kHz in Fig. 4a and Fig. 4b. The associated response and sensitivity under different bias magnetic fields are illustrated in Fig. 4a and Fig. 4c. In accordance with Eq. 5 and Eq. 6, the variation of the sensitivity results from the changes of the magnetostrictive coefficient $α$ of the Terfenol-D MR and the $Q_m$ of the Terfenol-D MR. Specifically, as shown in Fig. 4d, when the bias magnetic field increases from 0 Gs to 50 Gs, $α$ increases gradually and then reaches the saturation afterwards. However, the $Q_m$ decreases rapidly from 146 to 9 as the bias magnetic field increases from 0 Gs to 100 Gs and becomes constant when the bias magnetic field increases further. In fact, the resonant frequency of the fiber MR could be also changed by the bias magnetic field. Fig. 4b shows that the resonant frequency of fundamental flexural mode of the fiber MR increases by 5.63 kHz as the bias magnetic field increases from 0 Gs to 200 Gs. However, taking account of the practical performance, we exploit the mechanical resonance of the Terfenol-D MR, rather than that of the fiber MR.

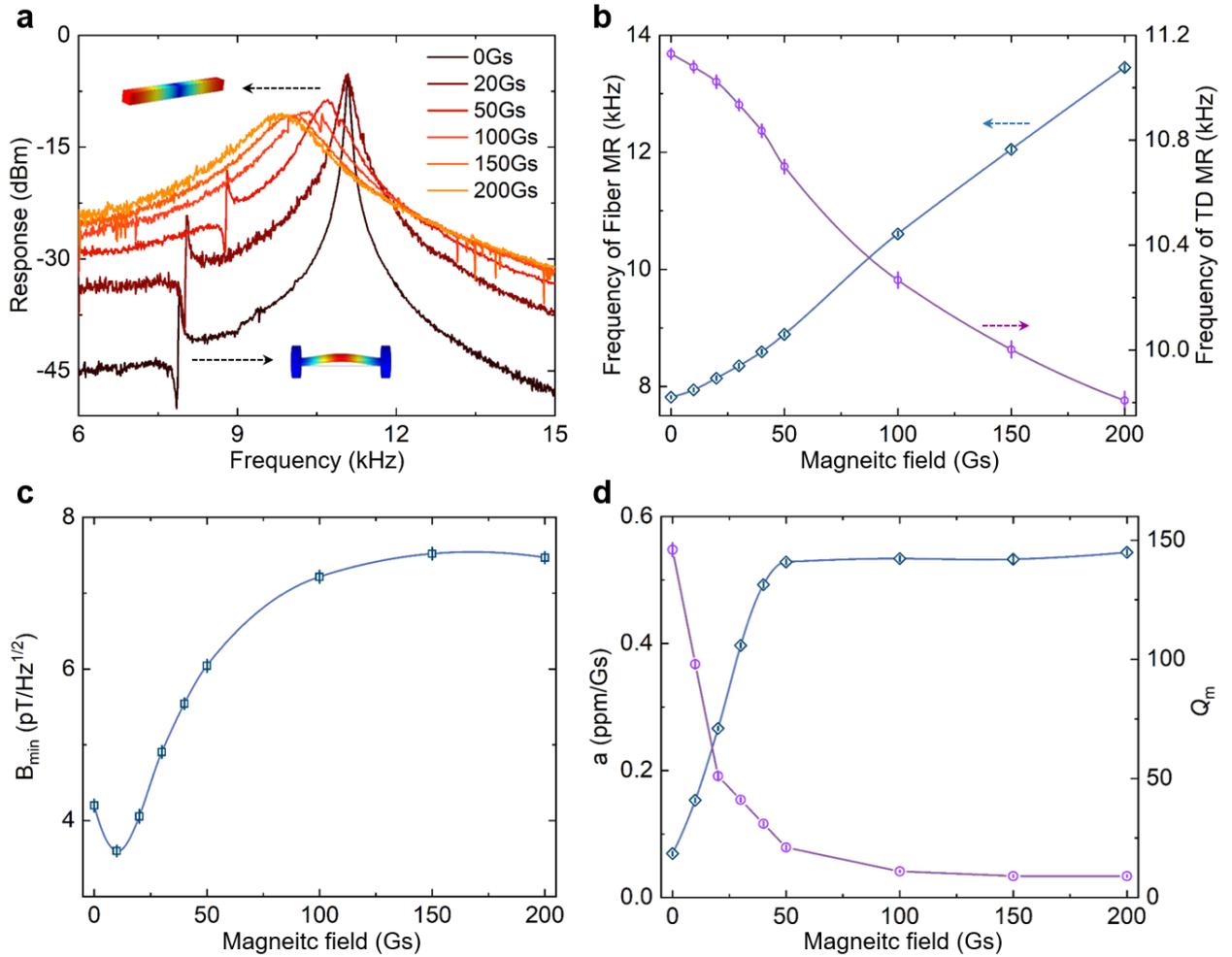

**Fig. 4. Tunable resonant responses. a**, Responses of the FOMM under different bias magnetic fields. **b** Resonant frequency shifts of the Terfenol-D and fiber MRs as a function of bias magnetic fields. **c**, AC sensitivity $B_{min}$ under different bias magnetic fields. **d**, $α$ and $Q_m$ as a function of bias magnetic fields. Inset: Finite-element modeling of mechanical resonant eigenmodes of the Terfenol-D and fiber MRs.

Finally, we evaluate the performances of the six FOMMs in our sensor network and calibrate their DC and AC magnetic-field sensitivities as summarized in Table 1. The DC magnetic-field sensitivity $B_{DC}$ is directly proportional to the $D_{TR}$, and is unrelated to the FWHM. The best sensitivity is 8.73 pm/Gs for the FOMM with a $D_{TR}$ of 13.5, and is slightly better than the best value of 7.57 pm/Gs for reported microcavity magnetometers[47], which is not applicable to AC magnetic-field measurement due to slow response of magnetic fluid. Notice that the measurable DC magnetic-field range of the FOMM based on Terfenol-D material is ten times larger than that of the microcavity magnetometers using magnetic fluid. This is attributed to the fact that the saturated magnetic intensities of the Terfenol-D material and magnetic fluid are about 2300 Gs and 220 Gs, respectively[50]. On the other hand, the AC magnetic-field sensitivity is proportional to the $D_{TR}$, and is inversely proportional to the FWHM of the PFBG. The best AC sensitivity of 537 fT/Hz$^{1/2}$ is achieved for the FOMM with FWHM of 1.4 pm and $D_{TR}$ of 5.1, and the best AC sensitivity of previous optomechanical magnetometer is 26 pT/Hz$^{1/2}$, where the DC response has not been presented[40].

The sensitivity of this magnetometer is mainly determined by the magnetostrictive mechanical resonator, PFBG, and deformation transfer radio. By using a mechanical resonator with larger magnetostrictive coefficient and increasing the deformation transfer radio, both DC and AC sensitivities can be improved. In addition, the AC sensitivity could be further enhanced by decreasing the FWHM of the PFBG and increasing the $Q_m$ of the magnetostrictive mechanical resonator.

**Table 1. Performances of the six FOMMs.** $D_{TR}$, deformation transferring radio; FWHM, full width at half maximum of PFBGs; $R_{max}$ is the maximum reflectivity of the PFBG; $R_{min}$ is the reflectivity of the PFBG at phase-shift point. $B_{DC}$, DC magnetic-field sensitivity; $B_{min}$, AC magnetic-field sensitivity.

| Sensors | $D_{TR}$ | FWHM (pm) | $R_{max}$ | $R_{min}$ | $B_{DC}$ (pm/Gs) | $B_{min}$ (pT/Hz$^{1/2}$) |
|---|---|---|---|---|---|---|
| 1 | 1 | 15.3 | 0.818 | 0.058 | 0.65 | 10.31 |
| 2 | 5.1 | 1.4 | 0.901 | 0.064 | 3.31 | 0.537 |
| 3 | 8.3 | 15.6 | 0.720 | 0.050 | 5.33 | 1.77 |
| 4 | 9.4 | 37.1 | 0.876 | 0.145 | 6.07 | 1.73 |
| 5 | 9.3 | 51.5 | 0.782 | 0.047 | 6.02 | 3.47 |
| 6 | 13.5 | 36.6 | 0.862 | 0.072 | 8.73 | 1.48 |

## Discussion

We report a unique cavity optomechanical sensor network, integrating the advantages of optomechanical ultrasensitive sensing mechanisms and advanced optical fiber sensing technologies. This optomechanical system enables robust low-noise low-loss coupling with light sources, offering invaluable opportunities to apply cavity optomechanical technologies in real-world applications. Moreover, FBGs are insensitive to polarization fluctuations and compatible with mature optical fiber multiplexing technologies, providing a promising candidate for long-distance sensing networks.

As a proof-of-concept, we demonstrate a one-dimensional directional optomechanical magnetometer network, achieving unprecedented DC and AC sensitivities at a magnetically unshielded ambient temperature and pressure. This network holds great potential for geomagnetic anomalies, space exploration missions, and mineral exploration. This sensing scheme could also be employed to measure strains, vibrations, and acoustic waves by choosing appropriate supports, offering promising candidates for the oil and gas industry, seafloor faults and ocean dynamics, and earthquake and volcano dynamic detection.

## Materials and Methods
### Fabrication of the FOMM

The key component of the proposed optomechanical sensor network is the fiber MR. The fiber MR is fabricated by packaging a standard PFBG into two separate silica capillaries with high-hardness adhesive or arc-discharge splicing progress[48]. By adjusting the distance between the two silica capillaries and the diameter of the PFBG, the resonant frequency of the fiber MR could be controlled accurately. In our experiments, the diameter of the PFBG is 125 μm, and the inner and outer diameters of the two silica capillaries are 126 and 1000 μm.

**Experimental setup for DC magnetic-field measurement**

To sense DC magnetic field, an amplified spontaneous emission (ASE) light source with a wavelength range from 1520 nm to 1570 nm is directed to the sensing network through a fiber circulator. The reflective light from the sensing network, which contains the deformation signals from the DC magnetic field, is monitored by an optical spectrum analyzer. By measuring the wavelength shifts of the PFBG, the DC magnetic field can be demodulated in real time, as shown in Fig. 2.

**Experimental setup for AC magnetic-field measurement**

The experimental setup for AC magnetic-field measurement is shown in Fig. 1. We use continuously tunable lasers (CTL) to monitor the deformation of the fiber MR, and demodulate the reflective light from the sensing network using balanced amplified photodetector, digital storage oscilloscopes, and electric spectrum analyzers. After obtaining the reflective spectrum of the PFBG, the wavelength of the probe light from the CTL is scanned around the resonant dip of the PFBG to choose a suitable value for the best slope $k$. The DC component from the photodetector is measured by the digital storage oscilloscope, and could be used to lock the laser to the PFBG. The corresponding AC component of the photodetector is monitored by the electric spectrum analyzer to achieve the power spectral densities of the noise and the response of the sensing network under AC magnetic-field excitation.

**Acknowledgments**

We thank Shangran Xie, Chang-Ling Zou, Beibei Li, and Jie Li for helpful discussions. This work is supported by: National Natural Science Foundation of China (grants no. 12174232, U21A6006, and 11804208) Innovation Program for Quantum Science and Technology (grants no. 2021ZD0300703 and 2023ZD0300400).